\documentclass[twocolumn,secnumarabic,amssymb, nobibnotes, aps, prl, superscriptaddress]{revtex4-1}

\usepackage{float}
\usepackage{graphicx}
\usepackage{amsmath} 
\usepackage{setspace}
\usepackage[shortlabels]{enumitem}
\usepackage{mathrsfs}
\usepackage{color,soul}
\usepackage[dvipsnames]{xcolor}
\usepackage{dcolumn,amssymb,subfig}
\usepackage{comment}
\usepackage[normalem]{ulem}

\usepackage{dcolumn} 
\usepackage{natbib}
\usepackage{bm} 
\usepackage{float} 
\usepackage{bbm} 
\usepackage{amsfonts}
\usepackage{lmodern} 
\usepackage{amsmath,amssymb}
\usepackage{bbold}

\renewcommand{\vec}[1]{\boldsymbol{#1}}
\newcommand{\tens}[1]{\boldsymbol{#1}}

\newcommand{\abs}[1]{\left| #1 \right|}

\usepackage{hyperref}
\hypersetup{
     colorlinks=true,
    urlcolor=blue,
    linkcolor=black,
    citecolor=black,
}
\begin{document}

\title{Active forces in confluent cell monolayers}
\author{Guanming Zhang}
\author{Julia M. Yeomans}
\affiliation{The Rudolf Peierls Centre for Theoretical Physics, Department of Physics, University of Oxford, Parks Road, Oxford OX1 3PU, UK}
\begin{abstract}
We use a computational phase-field model together with analytical analysis to study how inter-cellular active forces can mediate individual cell morphology and collective motion in a confluent cell monolayer. Contractile inter-cellular interactions lead to cell elongation, nematic ordering and active turbulence, characterised by motile topological defects.  Extensile interactions result in frustration, and perpendicular cell orientations become more prevalent. Furthermore, we show that contractile behaviour can change to extensile behaviour if anisotropic fluctuations in cell shape are considered.
\end{abstract}
\maketitle

\label{sec_Intro}

{\bf Introduction:} Asking how cells move collectively is a fascinating and important problem that encompasses both the concepts of forces and flows traditional to physics \cite{trepat2009physical,ladoux2017mechanobiology} and the molecular signalling which drives many biological phenomena \cite{boocock2021theory}. Generic descriptions of cell motility, such as the phase-field model \cite{mueller2019emergence,peyret2019sustained, zhang2020active,lober2014modeling,lober2015collisions,wenzel2019topological,loewe2020solid,palmieri2015multiple} and vertex models \cite{farhadifar2007influence,bi2015density,giavazzi2018flocking,bi2016motility,alt2017vertex}, have recently contributed to understanding several aspects of cell motility. 

\begin{figure}[b]
    \centering
    \includegraphics[width=0.5\textwidth]{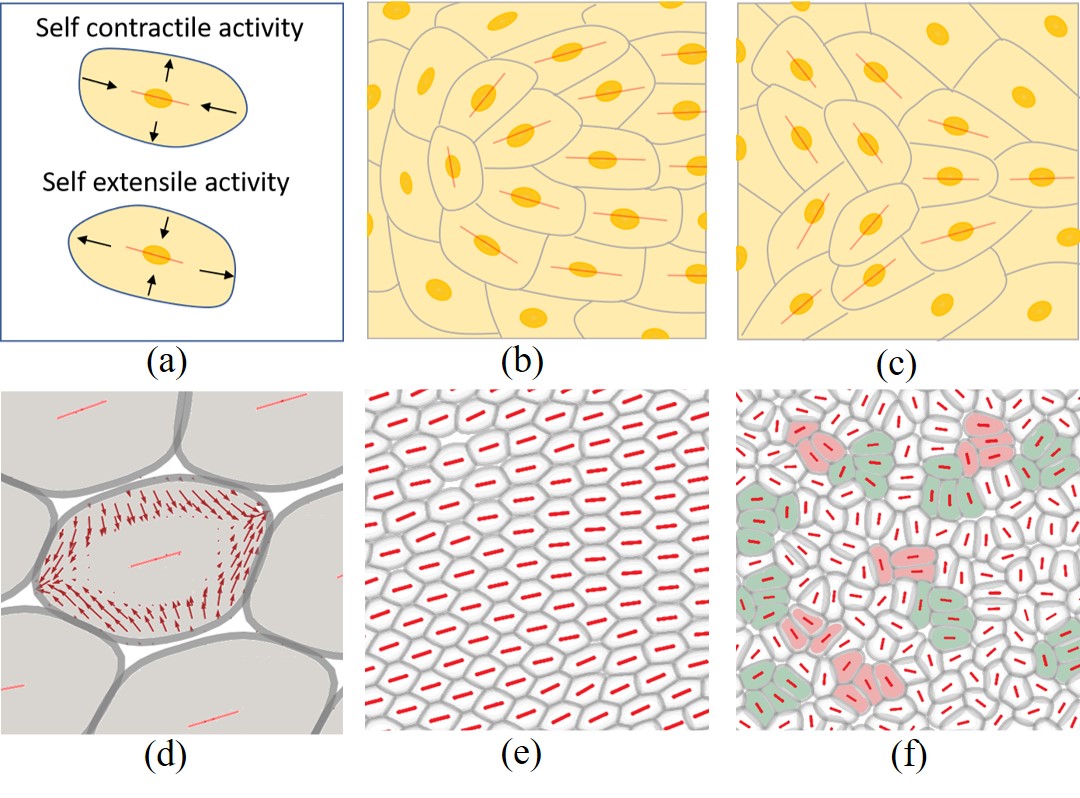}
    \caption{(a) self-deformation activity (b) schematic of +1/2 defect (c) schematic of -1/2  defect  (d) inter-cellular contractile forces act on the central cell as a self-extensile force (e) nematic ordering in tissue with contractile inter-cellular forces (f) `capped line' structures in tissue with extensile inter-cellular forces. Red bars indicate cell elongation axes.}
    \label{fig:schematics}
\end{figure}

The forces driving single cells across a flat surface are well understood. The cell is controlled by directional actin filaments, which can continuously polymerize and depolymerize to produce lamellopodia, protrusions that push the cell forwards \cite{alberts2003molecular,mitchison1996actin}. To advance, the cell needs to push against the substrate and to do this effectively it creates focal adhesions, which are mechanical links between internal actin bundles and the external surface \cite{sarangi2016coordination}. As it moves, the cell tends to polarize and elongate in the direction of motion \cite{parsons2010cell}. Contractile forces, mediated by myosin motors interacting with the actin network within the cell, tend to restore it to circular \cite{pollard2003cellular}. Thus a minimal physical model of single cell motility comprises a net force in the direction of the cell polarity, together with contractile, balanced forces restoring the cell to a circular shape. (We will refer to forces that tend to return an elongated cell to circular, or to extend it further, as {\em contractile} and {\em extensile} respectively, Fig.~\ref{fig:schematics}(a)).

Much less is understood about the dynamics of confluent cell layers. The cells can be jammed with local fluctuations \cite{bi2015density,sadati2013collective} or form liquid-like states where the motion has localised, correlated bursts of velocity or vorticity. Moreover, motile topological defects, regions where the long axes of the cells take comet or trefoil-like configurations (Fig.\ref{fig:schematics}(b)(c)), have been identified in several confluent cell layers \cite{blanch2018turbulent,balasubramaniam2021investigating,saw2017topological,kawaguchi2017topological}. This is reminiscent of active turbulence, which is the dynamical behaviour of many active nematic materials, such as suspensions of microswimmers \cite{sanchez2012spontaneous,dombrowski2004self} or microtubules  driven by motor proteins \cite{sanchez2012spontaneous}. However, the appearance of active turbulence requires elongated particles \cite{doostmohammadi2018active} and therefore it is somewhat surprising to identify topological defects even in assemblies of e.g.~MDCK cells that are on average isotropic in shape \cite{saw2017topological}. Moreover the comet defects can move towards their head, corresponding to extensile driving \cite{giomi2014defect}, even though individual cells are contractile \cite{balasubramaniam2021investigating}. Indeed, experiments and simulations have shown that the defect motion changes direction - indicating a change from extensile to contractile behaviour - as the cell-cell adhesion is varied. Other theoretical work has shown that fluctuating polar (unbalanced) forces can result in extensile defects \cite{patelli2019understanding, killeen2021polar}.

These observations raise questions about the identity of the physical forces governing collective cell motility.
The formation of lamellopodia is suppressed in confluent cell layers, a phenomenon termed contact inhibition of locomotion, suggesting the absence of any persistent, unbalanced (polar) forces \cite{abercrombie1970contact,stramer2017mechanisms}.  Therefore balanced forces must be acting to drive the cellular dynamics and, because the motion is persistent, these must be active, i.e. continuously fuelled by chemical energy. 
The most likely physical origin for these are the contractile forces within a cell which act through
inter-cellular junctions to pull on the cell's neighbours. 

We present analytical arguments and simulations, based on a two-dimensional, coarse-grained, phase-field model of cell motility, to show that active, contractile interactions between cells, mediated through cell junctions, lead to the cells elongating and lining up to give nematic ordering.  Decreasing cell-cell adhesion leads to  flows which destabilise the nematic order, resulting in active turbulence and contractile topological defects. 
We further show that anisotropic fluctuations of the inter-cellular forces can change the direction in which the defects move.

\noindent
{\bf Model:} The phase-field approach describing the dynamics of a confluent cell layer resolves individual cells and their interactions but not the internal cell machinery \cite{mueller2019emergence,lober2014modeling,ziebert2011model,ziebert2013effects}.
Each cell $i$ is represented by an individual phase field, $\varphi^{(i)}(\mathbf{x})$. The motion of each phase field is governed by a local velocity field, $\mathbf{v}^{(i)}(\mathbf{x})$, according to the equation of motion
\begin{equation}
\label{eq:dynamics}
\partial_t \varphi^{(i)}(\mathbf{x}) + \mathbf{v}^{(i)} (\mathbf{x}) \cdot \nabla \varphi^{(i)}(\mathbf{x} ) = -J_0 \frac{\delta {\mathcal{F}}}{\delta \varphi^{(i)} (\mathbf{x})}
\end{equation}
where $\mathcal{F}$ is a free energy.  $-J_0\frac{\delta {\mathcal{F}}}{\delta \varphi^{(i)}}$ describes the relaxation dynamics of the cells to a free energy minimum at a rate $J_0$. 
Assuming over-damped dynamics, the velocity $\mathbf{v}^{(i)} (\mathbf{x})$ of a cell is determined by the local force density acting on the cell,
\begin{equation}
\label{eq:force_balance}
\xi \mathbf{v}^{(i)} (\mathbf{x}) =\mathbf{f}_{\text{passive}}^{(i)} (\mathbf{x}) + \mathbf{f}_{\text{active}}^{(i)} (\mathbf{x}),
\end{equation}
where $\xi$ is a friction coefficient.

The passive force density,  $\mathbf{f}_{\text{passive}}^{(i)}(\mathbf{x}) = \frac{\delta \mathcal{F}}{\delta \varphi^{(i)}} \nabla \varphi^{(i)}$, includes a Cahn-Hilliard term that  encourages $\varphi^{(i)}$ to take values $1$, which we choose to correspond to the inside of the cell $i$, or $0$, which denotes the region outside the cell, a soft constraint, restricting the area of each cell, a repulsion energy that penalises overlap between cells and, of particular relevance here, a cell-cell adhesion energy with strength parameterised by $\omega$. 
See the SM \cite{SM} and
\cite{mueller2019emergence,zhang2020active,cates2018Theories,lober2015collisions} for more details.

To formulate the active contribution to the force density, $\mathbf{f}_{\text{active}}^{(i)} (\mathbf{x})$, we first calculate the deformation tensor that quantifies the shape of a cell \cite{mueller2019emergence,bigun1987optimal}, 
\begin{align}
\label{eq:deformation_tensor}
\begin{split}
{\mathcal D}^{(i)}
&= - \int d\mathbf{x} \left\{\nabla \varphi^{(i)} \nabla \varphi^{(i)^T} - {\frac{\mathbb{1}}{2}} \mbox{Tr}( \nabla \varphi^{(i)} \nabla \varphi^{(i)^T} )\right\} \\
& \equiv \sqrt{({\mathcal D}^{(i)}_{xx})^2 + ({\mathcal D}^{(i)}_{xy})^2}\; (\mathbf{d}_{\parallel}^{(i)} \mathbf{d}_{\parallel}^{(i)^T} -\mathbf{d}_{\bot}^{(i)} \mathbf{d}_{\bot}^{(i)^T}),\\
\end{split}
\end{align}
where $\mathbf{d}_{\parallel}^{(i)}$ and $\mathbf{d}_{\bot}^{(i)}$ are the orthonormal eigenvectors of the ${\mathcal D}^{(i)}$, along and perpendicular to the elongation axis of the cell respectively, normalised so that 
$\mathbf{d}_{\parallel}^{(i)}\mathbf{d}_{\parallel}^{(i)^T} + \mathbf{d}_{\bot}^{(i)}\mathbf{d}_{\bot}^{(i)^T} = \mathbb{1}$.


We next define a director, $ \vec{n}^{(i)}$, associated with each cell $i$ and
assume that $\vec{n}^{(i)}$ relaxes towards  $\mathbf{d}_{\parallel}^{(i)}$ 
through a stochastic relaxation process,
\begin{equation}
    \frac{d \mathbf{n}^{(i)}}{dt} = J_n(\mathbf{d}_{\parallel}^{(i)} - \mathbf{n}^{(i)}) + \boldsymbol{ \eta }^{(i)}(t).
    \label{eq:flucuation-dynamics}
\end{equation}
$J_n$ controls the time scale of relaxation. We assume anisotropic, Gaussian noise, uncorrelated between cells, with
\begin{align}
  \begin{split}
    &\langle \boldsymbol{\eta}(t) \rangle = \mathbf{0},\\
    &\langle \boldsymbol{\eta}^{(i)}(t) \boldsymbol{\eta}^{(i)}(t')^T \rangle = \\
    &\quad \quad  \delta(t-t')(D_{\parallel} \mathbf{d}_{\parallel}^{(i)}\mathbf{d}_{\parallel}^{(i)^T} + D_{\bot} \mathbf{d}_{\bot}^{(i)}\mathbf{d}_{\bot}^{(i)^T}).
  \end{split}
    \label{eq:noise}
\end{align}
The variance of the noise couples to the shape of the cell and can take different values for fluctuations along $\mathbf{d}_{\parallel}^{(i)}$ or 
 $\mathbf{d}_{\perp}^{(i)}$. 


 In the absence of any unbalanced active forces, the leading order contribution to the active stress acting on cell $i$ is related to the director field 
 by \cite{simha2002hydrodynamic}
\begin{equation}
\sigma_{\alpha \beta}^{(i)}(\mathbf{x})
= -\zeta_{\text{self}} \;  Q_{\alpha \beta}^{(i)}\; \varphi^{(i)}(\mathbf{x})
-\zeta_{\text{inter}} \sum_{ j \neq i} Q_{\alpha \beta}^{(j)} \;\varphi^{(j)}(\mathbf{x}) 
\label{eq:inter-stress}
\end{equation}
where
\begin{equation}
Q_{\alpha \beta}^{(i)} 
= \{(\vec{n}^{(i)}_\alpha \vec{n}^{(i)}_\beta - \frac{\abs{\vec{n}^{(i)}}^2}{2}  \delta_{\alpha \beta}\} .
\label{eq:Q-tensor}
\end{equation}
We distinguish between the stress acting on cell $i$ due to internal forces, of strength $\zeta_{\text{self}}$, and that due to other cells, of strength $\zeta_{\text{inter}}$. Our arguments are not changed qualitatively by the value of 
$\zeta_{\text{self}}$ and therefore we choose it to be zero.
The force density arising from the active stress is then
\begin{equation}
\mathbf{f}_{\text{active}}^{(i)}(\mathbf{x}) = \nabla \cdot \tens{\sigma}^{(i)} 
   =  -\zeta_\text{{inter}} \sum_{ j \neq i}  \mathbf{Q}^{(j)} \cdot \nabla \varphi^{(j)}(\mathbf{x}).
   \label{eq:active-force}
\end{equation}

\begin{figure*}[t]
    \centering
    \includegraphics[width=0.8\textwidth]{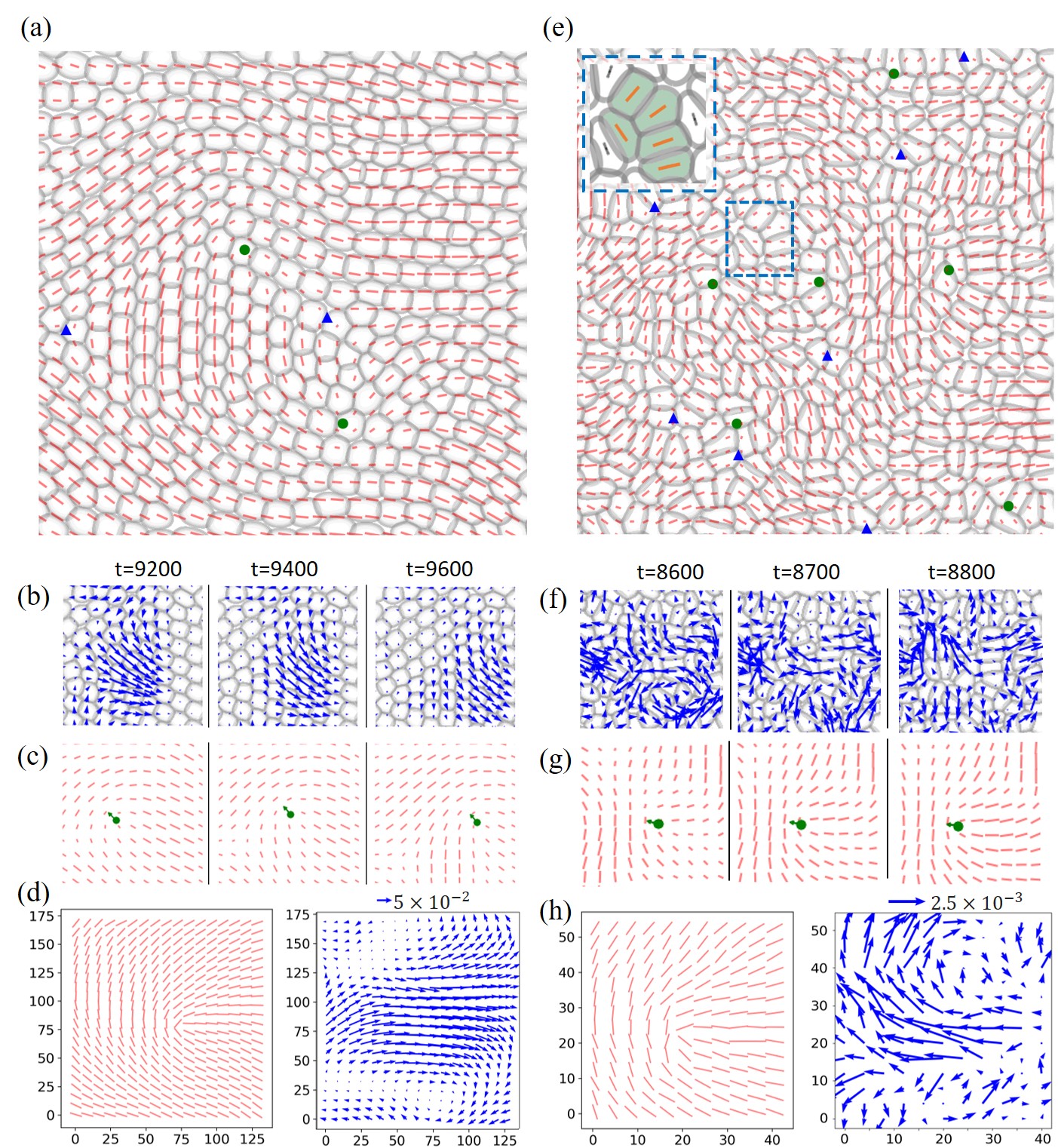}
    \caption{{\bf Cell and defect dynamics driven by inter-cellular forces.} Left column: contractile inter-cellular forces,  $\zeta_{\text{inter}}=-0.6$, $\omega=0$. Right column:  extensile inter-cellular forces,  $\zeta_{\text{inter}}=+0.4$, $\omega=0$.   
    (a),(e) Simulation snapshots.    Red lines indicates the coarse-grained field of elongation axes. Green dots and blue triangles indicate +1/2, comet-like, and -1/2, trefoil-like, defects. The sub-frame of (e) shows the `capped line' cell configuration.  
    (b),(f) Cell configuration around a $+1/2$ defect with imposed velocity field (blue arrows) at successive times.
    (c),(g) Corresponding images showing the elongation axes field, and the defect motion. The defect moves towards its tail (head) in the contractile (extensile) system. 
    (d),(h) Averaged cell elongation axes (left) and velocity (right) fields in the vicinity of a defect. Note the scale change between the contractile and extensile cases. Similar results for -1/2 defects are given in Fig.~S1.
 }
    \label{fig:no-fluc}
\end{figure*}

\noindent
{\bf Inter-cellular contractile forces:} 
Individual cells are contractile, so a plausible physical picture is that a cell feels the contractile forces from its neighbours, transmitted through cell-cell junctions.  Therefore, we first  investigate contractile, inter-cellular forces ($\zeta_{\text{inter}}<0$), assuming instantaneous relaxation of the director to the elongation axis of the cell $(\mathbf{n}^{(i)}=\mathbf{d}_{\parallel}^{(i)})$. 
Fig.~\ref{fig:schematics}(d) shows the typical  force density acting on a given cell due to its contractile neighbours. Surprisingly, the cell is {\em stretched}. 

This can be explained by considering Eq.~(\ref{eq:active-force}).
The gradient of a phase-field, $\nabla \varphi^{(i)}$, points perpendicular to a cell boundary towards the cell centre. In addition, in a confluent monolayer with strong cell-cell adhesion, cells nestle closely sharing common interfaces. These properties allow us to approximate $\nabla \varphi^{(j)} \approx -\nabla \varphi^{(i)}$ for the dominant contributions in Eq.~(\ref{eq:active-force}), so the active inter-cellular force can be written
\begin{equation}
\mathbf{f}_{\text{active}}^{(i)}(\mathbf{x}) 
= +\zeta_{\text{inter}} \ \mathbf{Q}^{(i)}_{\text{eff}} \cdot \nabla \varphi^{(i)}(\mathbf{x})
\label{eq:effectiveforcedensity}
\end{equation}
in terms of an effective $\mathbf{Q}$-tensor, $\mathbf{Q}^{(i)}_{\text{eff}} = \sum_{j \neq i } \mathbf{Q}^{(j)}$.
The change in sign shows that a cell with contractile neighbours will be subject to an extensile-like, self-deformation force density (Fig.~\ref{fig:schematics}(a)) which stretches the cell (Fig.~\ref{fig:schematics}(d)). Moreover, Eq.~(\ref{eq:effectiveforcedensity}) shows that each cell tends to align
along the averaged elongation direction of its neighbours which, following the usual mean-field argument, is expected to result in nematic ordering.

We check and extend this argument by solving the phase-field model numerically for $360$ cells (see SM for details \cite{SM}). 
For strong cell-cell adhesion, $\omega=0.5$, the cells are extended and aligned and become jammed in the configuration shown in {Fig~\ref{fig:schematics}(e)}.
Decreasing the cell-cell adhesion to zero weakens the alignment and allows the cells to move collectively. We observe the well-known splay instability that characterises contractile active nematics, and which leads to active turbulence.  Topological defects are continually created and destroyed. These move towards their tail confirming that they result from contractile forces {(Fig.~\ref{fig:no-fluc}(a)-(d), Movie 1).}


\noindent
{\bf Inter-cellular extensile forces:}
Although a contractile, inter-cellular force is more physical, it is interesting to compare extensile forcing ($\zeta_{\text{inter}}>0$). In this case, the alignment due to effective self-interactions is frustrated and there is no long-range ordering. Instead cells tend to align locally, at the scale of a few cells, in a {\em capped line} structure {(Fig.~\ref{fig:schematics}(f), Fig.~\ref{fig:no-fluc}(e), Movie 2)}.   Measuring the average velocity field around defects confirms that they move tail to head as expected in an extensile system ({Fig.~\ref{fig:no-fluc}(e)-(h)}).
Defects form more easily, and are more localised and less persistent than in the contractile case (note the different scales of {Fig.~\ref{fig:no-fluc}(d) and (h))}. 
It is of interest that similar capped line structures are seen in simulations of hard rods, driven by a polar force \cite{meacock2021bacteria,wensink2012meso}.

\noindent
{\bf Fluctuations} We next model the strong fluctuations in {cell shape} that are observed in many epithelial cell layers by including the anisotropic noise term, defined in Eq.\ (\ref{eq:noise}), in the equation governing the relaxation of a cell director, Eq.\ (\ref{eq:flucuation-dynamics}). This leads to anisotropic fluctuations of the inter-cellular forces. If the relaxation of the director to the long axis of  a cell is rapid compared to the time scale of cell reorientation,
Eq.~(\ref{eq:flucuation-dynamics}) can be integrated to give
\begin{equation}
    \mathbf{n}^{(i)} = \mathbf{n}^{(i)}_0 e^{-J_nt} + \mathbf{d}_{\parallel}^{(i)}(1-e^{-J_nt}) + \int_0^t ds \ \boldsymbol{\eta}(s)e^{-J_n(t-s)}
    \label{eq:expression-n}
\end{equation}
where $\mathbf{n}^{(i)}_0$ is the director of cell $i$ at $t=0$. Using this expression in Eqs.\ (\ref{eq:Q-tensor}) and (\ref{eq:active-force}) the averaged inter-cellular force follows as
\begin{eqnarray}
&&\langle \mathbf{f}^{(i)}_{\text{active}}(\mathbf{x}) \rangle
    \approx \nonumber \\
    &&-\zeta_{\text{inter}} (1 + \frac{D_{\parallel} - D_{\bot}}{2J_n})\sum_{j \neq i }(\mathbf{d}_{\parallel}^{(j)}\mathbf{d}_{\parallel}^{(j)^T} - \frac{1}{2}\mathbb{1}) \cdot \nabla \varphi^{(j)}(\mathbf{x}). \nonumber \\
    &&
  \label{mean-fint}
\end{eqnarray}
Hence fluctuations in {cell shape} along $\mathbf{d}_{\bot}$ tend to change the sign of the activity.

\begin{figure}[h]
    \captionsetup{justification=raggedright,singlelinecheck=false}
    \includegraphics[width=0.5\textwidth]{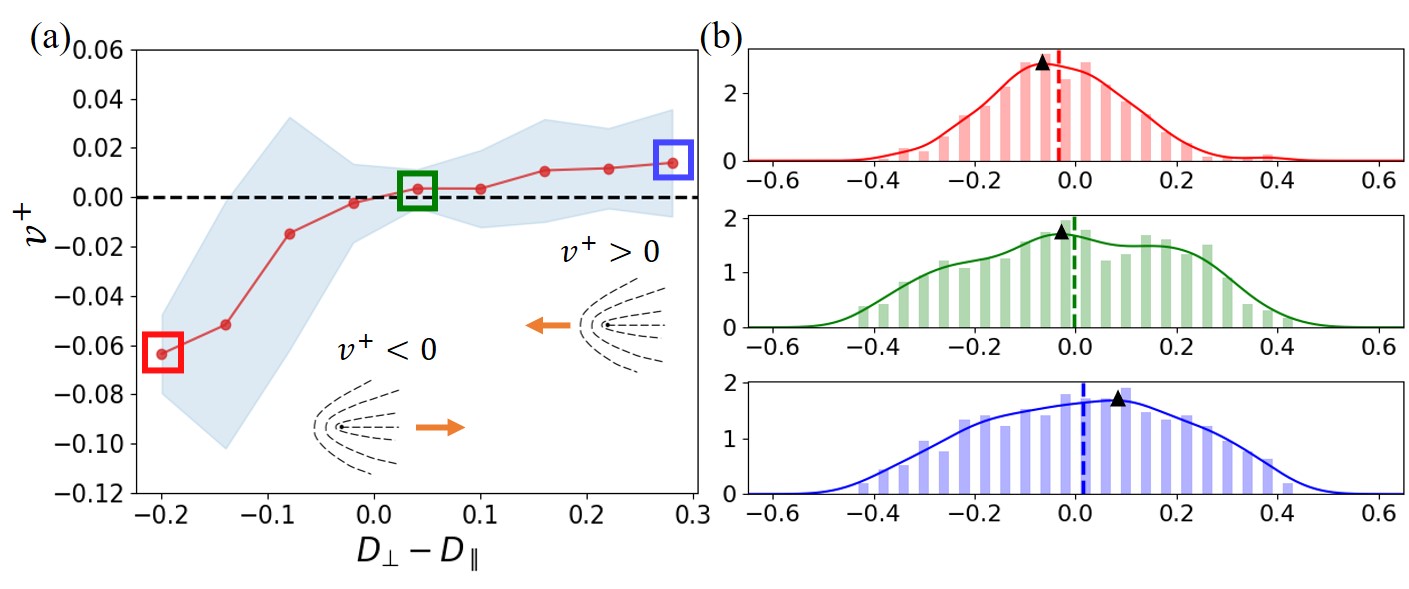}
    \caption{(a) Velocity of the +1/2 defects along their tail to head direction, $v^{+}$, as the magnitude of the fluctuations along the short axis of the cells, $D_{\bot}$, is varied. The red line is the average velocity and the light blue background denotes the standard error. 
    (b) Histogram of $v^{+}$corresponding to framed points in (a). Vertical dashed lines: mean value. Black triangles: maximum of the smoothed histogram.
    $ \zeta_{\text{inter}} = -0.6, D_{\parallel} = 0.2$.}
    \label{fig:plus-defect-velocity}
\end{figure}

This is confirmed in Fig.~\ref{fig:plus-defect-velocity} where we plot the defect velocity against $D_{\bot}-D_{\parallel}$, showing a change from contractile to extensile behaviour (see also Fig.~S2, Movies 3,4). 
Further evidence for a crossover is presented in Fig.~S3 where the distribution of the angles between the long axes of neighbouring cells is plotted. With increasing $D_{\bot}$ there is a crossover from a peak at $0^o$ signalling nematic ordering  to two weaker peaks at  $0^o$ and  $90^o$ indicating the capped line state.
When extensile and contractile influences balance (green square in Fig.~\ref{fig:plus-defect-velocity}), a jammed phase emerges where the shape of cells is round on average, but fluctuates strongly (Movie 5).\\

Our work gives an explanation for why cells that are, on average, circular can exhibit nematic ordering and active turbulence in terms of fundamental, active, interactions between the cells. It also shows that the direction of motion of active defects can vary. Our prediction that this depends on anisotropic fluctuations of inter-cellular forces and hence of cell shape could be tested experimentally by analysing the dynamical evolution of monolayers.

\end{document}


\title{Active forces in confluent cell monolayers}
\author{Guanming Zhang}
\author{Julia M. Yeomans}
\affiliation{The Rudolf Peierls Centre for Theoretical Physics, Department of Physics, University of Oxford, Parks Road, Oxford OX1 3PU, UK}
\maketitle

\begin{center}
    {\bf Supplementary Material}
\end{center}

\section{Movie description}
{\bf Movie 1}: Contractile confluent cell layer without fluctuations. $\zeta_{\text{inter}} = -0.6, \omega=0.0$. \\

{\bf Movie 2}: Extensile confluent cell layer without fluctuations. $\zeta_{\text{inter}} = +0.4, \omega=0.0$. \\

{\bf Movie 3}: Confluent cell layer with inter-cellular force fluctuations showing contractile behaviour.\\ $\zeta_{\text{inter}} = -0.4, \omega=0.0, D_{\parallel} = 0.2,D_{\bot} = 0.0$. \\

{\bf Movie 4}: Confluent cell layer with inter-cellular force fluctuations showing extensile behaviour.\\    $\zeta_{\text{inter}} = -0.4, \omega=0.0, D_{\parallel} = 0.2,D_{\bot} = 0.6$.\\

{\bf Movie 5}: Fluctuating jammed cells. $\zeta_{\text{inter}} = -0.4, \omega=0.0, D_{\parallel} = 0.2,D_{\bot} = 0.3$.\\

In the movies, the blue triangles and the green dots locate the positions of +1/2 and -1/2 defects respectively. Black bars at the centre of cells show the elongation axes. 

\section{Passive forces in the phase-field model}

We write the free energy as  $\mathcal{F} = \mathcal{F}_{\text{CH}} + \mathcal{F}_{\text{area}} + \mathcal{F}_{\text{rep}} + \mathcal{F}_{\text{adh}}$ \cite{mueller2019emergence,zhang2020active}.
The first contribution, defined as
\begin{equation}
{\mathcal F}_{\text{CH}}=\sum_i \frac{\gamma}{\lambda} \int d \mathbf{x} \left\{ 4 \varphi^{(i)}(\mathbf{x}) ^2(1-\varphi^{(i)}(\mathbf{x} ))^2+\lambda^2[\nabla \varphi^{(i)}(\mathbf{x} )]^2\right\},
\end{equation}
is a Cahn-Hilliard free energy that  encourages $\varphi^{(i)}$ to take values $1$ or $0$ inside or outside a cell. The cell boundary is located at $\varphi^{(i)} = 1/2$ and has width $\sim 2\lambda$, set by the gradient term, and ${\gamma}/{\lambda}$ is an energy scale. The Cahn-Hilliard free energy controls the deformability of the individual cells. 

In addition to the cell deformability, we account for the cell compressibility using a free energy term
\begin{equation}
{\mathcal F}_{\text{area}}=\sum_i \mu \left\{ 1-\frac{1}{\pi R^2} \int d \mathbf{x} \; \varphi^{(i)}(\mathbf{x})^2 \right \}^2,
\end{equation}
which imposes a soft constraint, of strength $\mu$, restricting the area of each cell to $\pi R^2$. Therefore, in the absence of any active forces or cell-cell interactions, the cells will relax to circles. 

The passive contributions to cell-cell interactions are introduced through repulsion and adhesion free energies as
\begin{equation}
{\mathcal F}_{\text{rep}}=\sum_i \sum_{j \neq i} \frac{\kappa}{\lambda} \int d \mathbf{x}\; \varphi^{(i)}(\mathbf{x})^2 \varphi^{(j)}(\mathbf{x})^2,
\end{equation}
which penalises overlap between cells with an energy scale ${\kappa}/{\lambda}$, and 
\begin{equation}
{\mathcal F}_{\text{adh}}=\sum_i \sum_{j \neq i} \omega\lambda \int d \mathbf{x}\; \nabla \varphi^{(i)} \cdot \nabla \varphi^{(j)},
\label{adhesion}
\end{equation}
which is a term favouring cell-cell adhesion. Note that $ \lambda \intg{\nabla \varphi^{(i)} \cdot \nabla \varphi^{(j)} }$ measures the length of the contact line between two cells and $\omega / \lambda$ is an energy scale.
If no active forces or external forces are acting, the cell layer will relax to a minimum of the total free energy. For cells in a confluent layer the global minimum is static, identical hexagons. 

The passive force reads \cite{cates2018Theories,mueller2019emergence,palmieri2015multiple}
 $$\mathbf{f}^{(i)}_{\text{passive}}(\mathbf{x}) = \frac{\delta \mathcal{F}}{\delta \varphi^{(i)}} \nabla \varphi^{(i)}.$$
In particular, the passive cell-cell adhesive force for cell $i$ follows from the free energy as
\begin{equation}
 \frac{\delta \mathcal{F}^{\text{adh}}}{\delta \varphi^{(i)}} \nabla \varphi^{(i)}  = \omega \lambda \nabla \varphi^{(i)} \sum_{j \neq i} \nabla^2 \varphi^{(j)}.
\end{equation}
However, to improve numerical stability, we constrain the $\nabla^2\varphi_i$ term by taking
 \begin{equation}
    \mathbf{f}^{(i)}_{\text{adh}}(\mathbf{x}) = \omega \lambda \nabla \varphi^{(i)} \sum_{j \neq i} g(\nabla^2 \varphi^{(j)})
    \text{, where } g(x) = \frac{x}{\sqrt{1 + x^2}}.
    \label{eq:adhesive-force}
\end{equation}
 g($\nabla^2 \varphi$) restrains the magnitude of $\nabla^2 \varphi$, and $ \mathbf{f}^{(i)}_{\text{adh}}(\mathbf{x}) $ approaches $\frac{\delta \mathcal{F}^{\text{adh}}}{\delta \varphi^{(i)}} \nabla \varphi^{(i)}$ with decreasing $\nabla^2 \varphi $. A similar formulation is used in \cite{lober2015collisions}. \\

\section{Simulation details}
The equations of the phase field dynamics are
solved using a finite difference method in space and a predictor-corrector method in time \cite{mueller2019emergence}.\\

We simulate the dynamics of $360$ cells of radius $R = 8$ in a periodic domain of size $280\times 280$ lattice sites. Initially, cells with radius $R/2$ are placed randomly, but with the constraint that the distances of neighbouring centres $> 0.8R$. They are then relaxed for $3000$ time steps under passive dynamics to reach  confluence. The simulations then start with a randomly oriented nematic director. They are run for $30000$ time steps and data is collected after $3000$ time steps. Parameter values are  $\lambda=2.0$, $\gamma=1.4$, $\kappa=1.5$, $\mu=120$, $\xi=3.0$ and $J_0=0.005$. We assume instantaneous relaxation of the director to the elongation axis of the cell ($\mathbf{n}^{(i)} = \mathbf{d}_{\parallel}^{(i)} $) for simulations without fluctuations. We use $J_n=0.1$ for a fast relaxation  scheme when solving Eq.(4) for fluctuating inter-cellular forces.\\ 

To include fluctuations, we obtain independent random numbers $x_1^{(i)}(t)$, $x_2^{(i)}(t)$  from Gaussian distributions with zero mean and unit variance at each time step $t$ and for each cell $i$. The anisotropic Gaussian noise is then calculated as $\boldsymbol{ \eta }^{(i)}(t) = \sqrt{D_{\parallel}} x_1^{(i)}(t) \mathbf{d}_{\parallel}^{(i)} + \sqrt{D_{\bot}} x_2^{(i)}(t) \mathbf{d}_{\bot}^{(i)}$. We use the Euler-Maruyama method to update the stochastic relaxation process described by Eq.(4).

\section{Velocity field, deformation tensor field and defect tracking} 
The centre-of-mass position and velocity of cell $i$ are defined as 
\begin{equation}
   \mathbf{x}_{com}^{(i)} = \frac{\int d \mathbf{x} \varphi^{(i)}(\mathbf{x}) \mathbf{x}}{\int d \mathbf{x} \varphi^{(i)}(\mathbf{x})},     
   \quad    
   \mathbf{v}_{com}^{(i)} = \frac{d \mathbf{x}^{(i)}_{com}}{dt} .
\end{equation}
The coarse-grained velocity field and deformation tensor field of the cell layer are defined by 
\begin{equation}
\mathbf{v}(\mathbf{x}) = \sum_{i} \varphi^{(i)} \mathbf{v}^{(i)}_{com}, \quad \mathcal{D}(\mathbf{x}) = \sum_{i} \varphi^{(i)}(\mathbf{x}) \mathcal{D}^{(i)},
\end{equation}
 %
where ${D}^{(i)}$ is the deformation tensor. 

Velocity fields and the field corresponding to the cell elongation axes around defects are calculated by smoothing the weighted velocity field and deformation tensor field over a sliding window of size 3R by 3R \cite{mueller2019emergence}. A defect is detected by examining the deformation tensor field of its 8 neighbouring points. The orientation of the defects is calculated by the method in \cite{vromans2016orientational}. Then, we crop and align the fields around each defect and average over all the defects for all time steps.

To obtain the +1/2 defect velocities, we store two lists of +1/2 defect positions at time step $t$ and $t + \Delta t$ as $\{\mathbf{x}_d^i(t)\}$ and $\{\mathbf{x}_d^j(t+\Delta t)\}$.Then, we pair each defect $p$ at time $t$ with its nearest defect $q$ at time $t+\Delta t$. Defects $p$ and $q$ are considered to be the same defect at different time steps if $| \mathbf{x}_d^q(t+\Delta t) - \mathbf{x}_d^p(t)| < v_{t} \Delta t$ where $v_{t}$ is a chosen threshold velocity and the defect velocity is calculated as $\mathbf{v}^p(t) = (\mathbf{x}_d^q(t+\Delta t) - \mathbf{x}_d^p(t))/\Delta t$. If $| \mathbf{x}_d^q(t+\Delta t) - \mathbf{x}_d^p(t)|  \geq v_{t} \Delta t$, defect $q$ is consider as a newly created defect at time $t+\Delta t$. We set $\Delta t = 30$, a number much smaller than the defect life-time ($\sim$ 200), and velocity threshold $v_t = 1.2R/30$. For simulations with additional force fluctuations, we filter out the defects whose lifetime is $\leq 60$ to eliminate those created by transient shape fluctuations.

The velocity of each defect is projected to the tail to head direction to get $v^+$. The mean projected defect velocities of 8 simulations with the same $D_{\bot}$ are used to obtain the average value and standard error corresponding to each data point in Fig. 3. Then we vary $D_{\bot}$ and repeat the steps to get the whole curve. In Fig. 3(b), each histogram shows the distribution of $v^{+}$ in one representative simulation.
\clearpage

\begin{center}
 {\bf Supplementary Figures}   
\end{center}
\begin{figure}[H]
    \centering
    \includegraphics[width=0.9\textwidth]{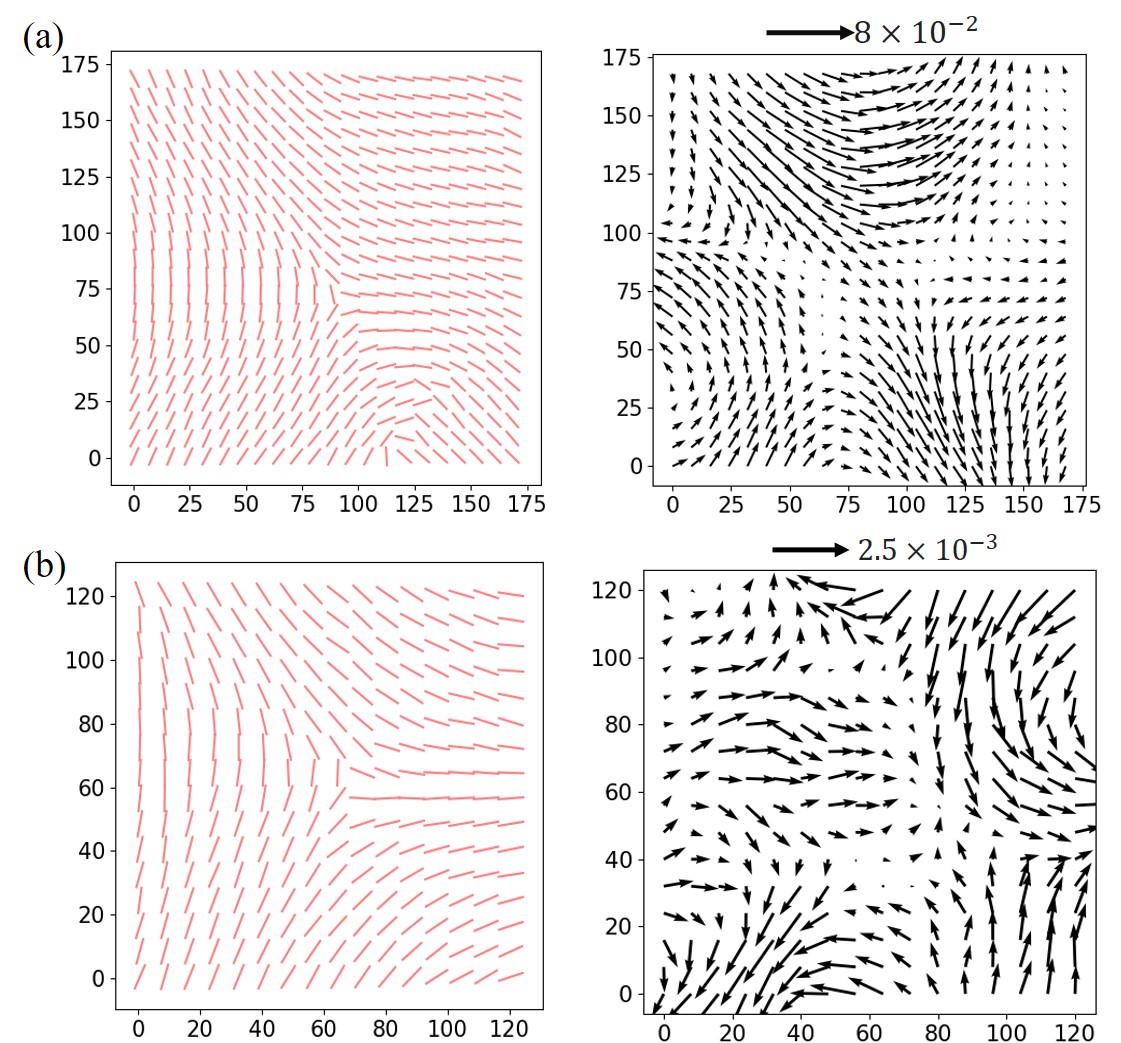}
    \caption{{ \bf Trefoil-like, -1/2, defects in simulations without fluctuations.} Averaged cell elongation axes (left) and velocity (right) fields in the vicinity of a $-1/2$ defect for (a) a contractile system with $\zeta_{\text{inter}}=-0.6$ (b) a extensile system  with $\zeta_{\text{inter}}=+0.4$. Note the scale change in the velocity between the contractile and extensile cases.  }
    \label{fig:-1/2-defects}
      
\end{figure}

\begin{figure}[H]
    \centering
    \includegraphics[width=0.7\textwidth]{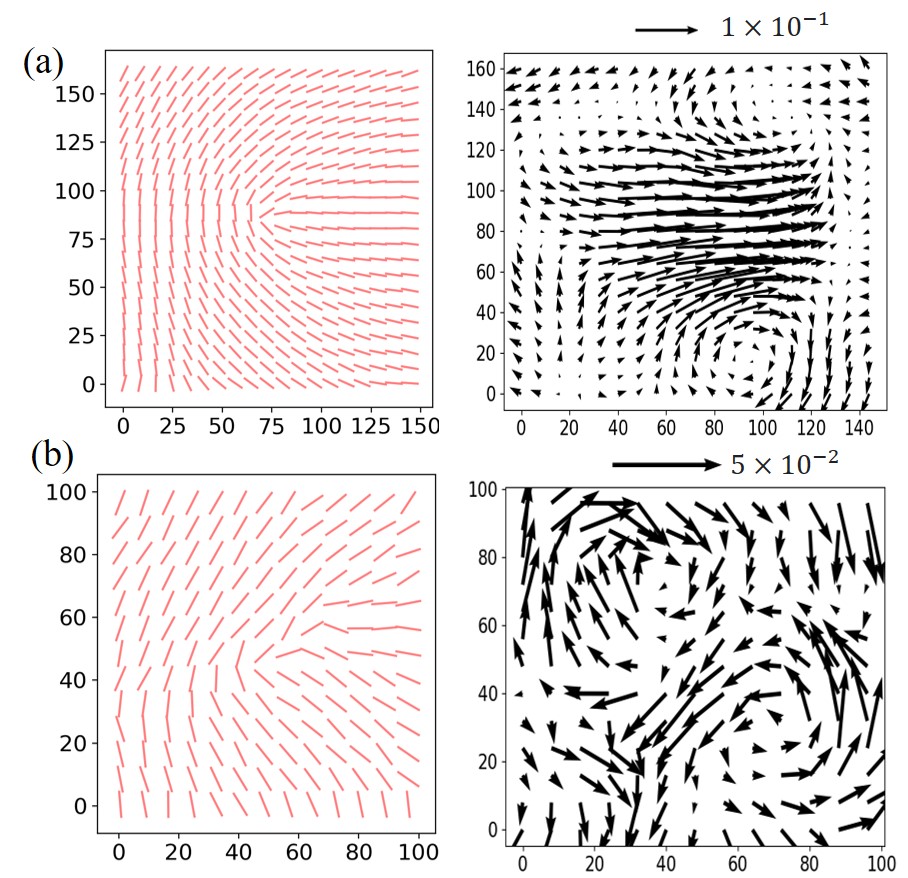}
    \caption{{\bf Behaviour of comet-like, $+1/2$, defects as anisotropic fluctuations are introduced.}
   Averaged cell elongation axes (left) and velocity (right) fields in the vicinity of a $+1/2$ defect for (a) a contractile system with no fluctuations (b) a contractile system with anisotropic fluctuations which result in extensile-like behaviour. $\zeta_{\text{inter}}=-0.4,D_{\parallel}=0.1, D_{\bot} = 0.6$. }
    \label{fig:fluc}
\end{figure}

\begin{figure}[H]
    \centering
    \includegraphics[width=0.6\textwidth]{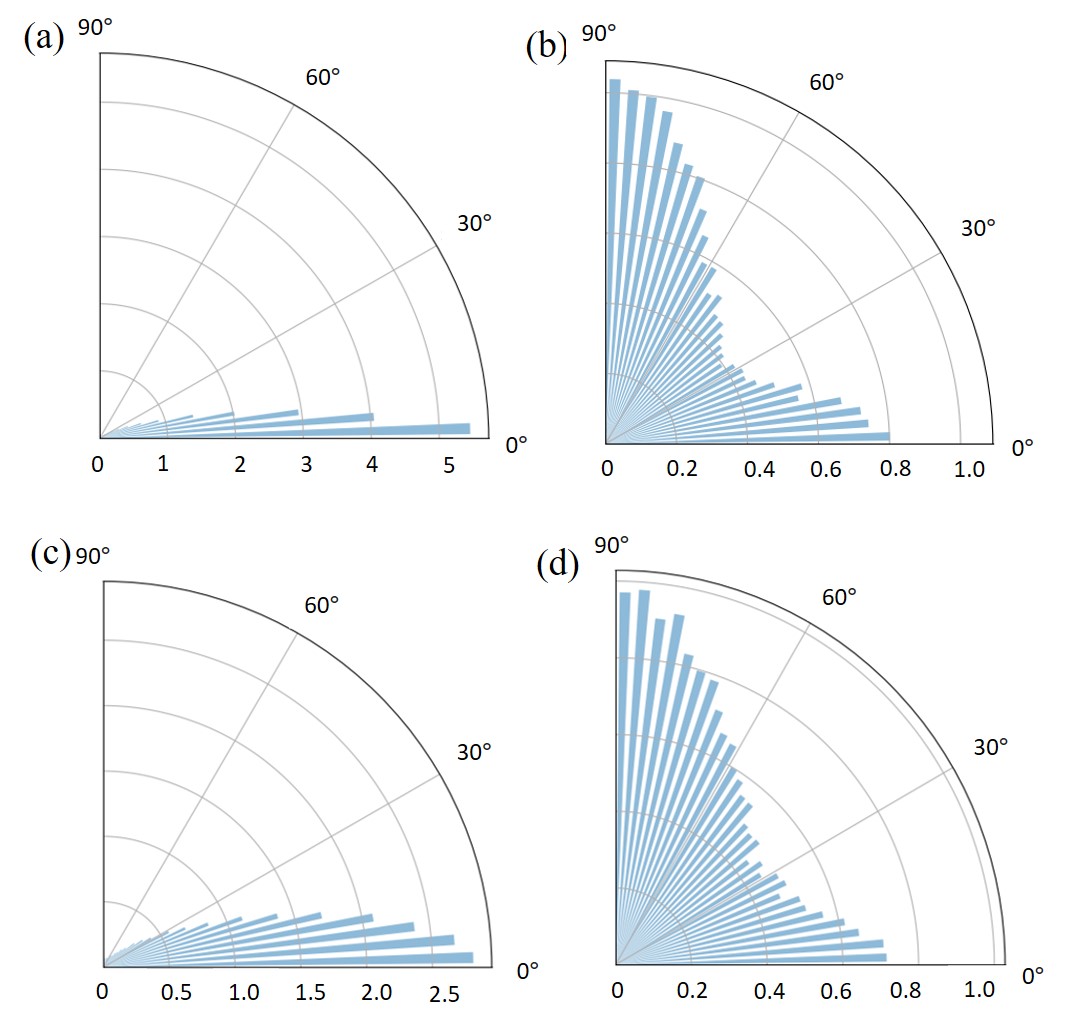}
    \caption{{\bf Distributions of the elongation axes of neighbouring cells.} (a) contractile forces, $\zeta_{\text{inter}}=-0.6$. (b) extensile forces, $\zeta_{\text{inter}}=+0.4$. (c) contractile forces with fluctuations, $\zeta_{\text{inter}}=-0.4, D_{\parallel} = 0.2,D_{\bot} = 0.0$. (d) contractile forces with anisotropic fluctuations that lead to effective extensile behaviour, $\zeta_{\text{inter}}=-0.4, D_{\parallel} = 0.0,D_{\bot} = 0.6$. In (a) and (c), most of the cells line up with their neighbours except for cells around defects. In (b) and (d), a "capped line" structure is favoured. Therefore, the distribution peaks at $90^{o}$ and $0^{o}$.
    }
    \label{fig:fluc}
\end{figure}

\clearpage